\documentclass[prd,superscriptaddress,amsfonts,amssymb,twocolumn,amsmath,showpacs]{revtex4-2}
\usepackage{bm}
\usepackage{amsfonts}
\usepackage{tensor}

\usepackage{latexsym}
\usepackage[utf8]{inputenc}
\usepackage{graphicx}
\usepackage{amsmath}
\usepackage{palatino}
\usepackage{mathpazo}
\usepackage[british]{babel}
\usepackage{hhline}
\usepackage{multirow}
\usepackage{textcomp}
\usepackage{float}
\usepackage{booktabs}
\usepackage{dcolumn}
\usepackage{lipsum} 
\usepackage{mdframed}
\usepackage[]{mdframed}
\usepackage{hhline}
\usepackage{multirow}
\usepackage{ragged2e}
\usepackage{hyperref}
\hypersetup{colorlinks,citecolor=blue}
\hypersetup{colorlinks=true,linkcolor=red,filecolor=magenta,    urlcolor=cyan}
\usepackage{amsmath}
\usepackage{xcolor}
\usepackage{orcidlink}
\usepackage{epsfig}
\usepackage{caption}
\usepackage{subcaption}
\usepackage{commath}

\def\jnl@style{\it}
\def\aaref@jnl#1{{\jnl@style#1}}

\def\aaref@jnl#1{{\jnl@style#1}}

\def\aj{\aaref@jnl{AJ}}                   
\def\apj{\aaref@jnl{ApJ}}                 
\def\apjl{\aaref@jnl{ApJ}}                
\def\apjs{\aaref@jnl{ApJS}}               
\def\apss{\aaref@jnl{Ap\&SS}}             
\def\aap{\aaref@jnl{A\&A}}                
\def\aapr{\aaref@jnl{A\&A~Rev.}}          
\def\aaps{\aaref@jnl{A\&AS}}              
\def\mnras{\aaref@jnl{Mon.~Not.~Roy.~Astron.~Soc.}}             
\def\prd{\aaref@jnl{Phys.~Rev.~D}}        
\def\prc{\aaref@jnl{Phys.~Rev.~C}}  
\def\prl{\aaref@jnl{Phys.~Rev.~Lett.}}    
\def\qjras{\aaref@jnl{QJRAS}}             
\def\skytel{\aaref@jnl{S\&T}}             
\def\ssr{\aaref@jnl{Space~Sci.~Rev.}}     
\def\zap{\aaref@jnl{ZAp}}                 
\def\nat{\aaref@jnl{Nature}}              
\def\aplett{\aaref@jnl{Astrophys.~Lett.}} 
\def\apspr{\aaref@jnl{Astrophys.~Space~Phys.~Res.}} 
\def\physrep{\aaref@jnl{Phys.~Rep.}}      
\def\physscr{\aaref@jnl{Phys.~Scr}}       
\def\commat{\aaref@jnl{Comm.~Math.~Phys.}}              
\def\science{\aaref@jnl{Science}}               
\def\cqg{\aaref@jnl{Classical Quant.~Grav.}}            
\def\jpcs{\aaref@jnl{JPCS}}                                     
\def\ijmpd{\aaref@jnl{Int.~J.~Mod.~Phys.~D}}                    
\def\grg{\aaref@jnl{Gen.~Relat.~Gravit.}}               
\def\rpp{\aaref@jnl{Rep.~Prog.~Phys.}}          
\def\npa{\aaref@jnl{Nucl.~Phys.~A}}        
\def\lrr{\aaref@jnl{Living Rev.~Rel.}}                   
\def\jcap{\aaref@jnl{J.~Cosmology Astropart.~Phys.}}    
\def\rmp{\aaref@jnl{Rev.~Mod.~Phys.}}   
\def\epjc{\aaref@jnl{Eur.~Phys.~J.~C}} 
\def\plb{\aaref@jnl{~Phy.~Lett.~B}} 
\def\mpla{\aaref@jnl{Mod.~Phy.~Lett.~A}} 
\def\arxiv{\aaref@jnl{arxiv.org}}

\allowdisplaybreaks[1]
\renewcommand{\arraystretch}{1.1}
\begin{document}

\title{Cosmological Dynamics on a Novel $f(Q)$ Gravity Model with Recent  DESI DR2 Observation}

\author{S. A. Kadam\orcidlink{0000-0002-2799-7870}}
\email{siddheshwar.kadam@dypiu.ac.in;
\\k.siddheshwar47@gmail.com}
\affiliation{Centre for Interdisciplinary Studies and Research, D Y Patil International University, Akurdi, Pune-411044, Maharashtra, India.}

\author{D. Revanth Kumar\orcidlink{0009-0002-6599-3608}}
\email{2406c9m003@sru.edu.in}
\affiliation{Department of Mathematics, SR University, Warangal-506371,
Telangana, India.}

\author{Santosh Kumar Yadav\orcidlink{0009-0009-2581-387X}}
\email{sky91bbaulko@gmail.com}
\affiliation{Department of Mathematics, SR University, Warangal-506371,
Telangana, India.}

\begin{abstract}
\textbf{Abstract:} In this article, we investigate the cosmological viability of a modified symmetric teleparallel gravity model within the \( f(Q) \) framework. We derive observational constraints on the model parameters by performing a Markov Chain Monte Carlo analysis using a combined dataset consisting of cosmic chronometers, PantheonPlus SH0ES, and DESI BAO DR2. Our analysis yields the best-fit values for the model parameters \( m=-0.386 \pm 0.090 \) and \( n=-1.055 \pm 0.047 \), along with the cosmological parameters at present: \( H_0 = 73.19 \pm 0.25 \), \( q_0 = -0.51 \pm 0.6 \), and \( \omega_{0} = -0.73 \pm 0.3 \), at 68\% CL. Furthermore, we examine the physical behavior of the model, focusing on the effective equation of state and deceleration parameter. Our findings indicate that the model experiences a transition from the early deceleration phase to the late-time cosmic acceleration, and the transition occurs at a redshift \( z_{tr} = 0.573\). We also analyse the \( om(z) \) diagnostic, which reflects a positive slope, supporting the behavior of the equation of state parameter in the quintessence region.
\end{abstract}

\maketitle

\section{Introduction}\label{Introduction}
The detection of the late-time accelerated expansion of the Universe, first uncovered through observations of Type Ia supernovae~\cite{Riess:1998cb, Perlmutter:1998np} and later validated by cosmic microwave background (CMB) anisotropy measurements~\cite{Caldwell:2004xf, Huang:2005re}, baryon acoustic oscillations (BAO)~\cite{SDSS:2005xqv, SDSS:2009ocz}, and large-scale structure surveys~\cite{Koivisto:2005mm, Daniel:2008et}, has greatly influenced modern cosmology. In the context of General Relativity (GR), this acceleration is commonly linked to dark energy (DE), usually represented as a cosmological constant $\Lambda$. While the $\Lambda$CDM model~\cite{Planck:2018vyg} provides an outstanding phenomenological account of a broad spectrum of observational datasets. Still, it faces ongoing theoretical complications, including the fine-tuning~\cite{Weinberg:1988cp, Carroll:1991mt} and the coincidence problem~\cite{Zlatev:1998tr}. These challenges have driven the exploration of modified gravity (MG) theories as alternative frameworks for explaining cosmic acceleration without depending on exotic fluid components. The interaction of gravity within spacetime can be characterized by three geometric features: curvature, torsion, and non-metricity~\cite{Di_Valentino_2025, Narawade:2025, Duchaniya:2024, Agrawal:2021, Samaddar:2025, Samaddar:2025b, Bahamonde:2019shr}. GR is fundamentally based on curvature, with both torsion and non-metricity being absent, where teleparallel gravity operates under the assumptions of zero curvature and non-metricity~\cite{bahamonde:2021teleparallel, LeviSaid:2021yat}. The foundation of symmetric teleparallel gravity assumes the absence of curvature and torsion~\cite{Albuquerque:2022eac, Solanki:2022ccf}. This phenomena arise from alterations in the lengths of vectors during parallel transport, contrasting with the directional changes~\cite{Xu:2019sbp}.

The $f(R)$ gravity represents an initial extension of GR~\cite{Starobinsky:2007,Sotiriou:2008rp}, while $f(T)$ gravity~\cite{Farrugia:2016xcw,Ferraro:2006jd,Bengochea:2008gz} modifies the teleparallel gravity. The $f(Q)$ gravity, which offers initial modification to symmetric teleparallel gravity, has been proposed~\cite{BeltranJimenez:2017tkd}. In the domain of Weyl-Cartan geometry, the symmetric metric tensor $g_{\mu\nu}$ is utilized to define vector lengths, with the covariant derivative and parallel transport being governed by an asymmetric affine connection $\tilde{\Gamma}^{\gamma}_{\mu\nu}$. An extension of symmetric teleparallel gravity defines the gravitational action \( L \) as a function \( f \) that depends on the non-metricity \( Q \) and the trace of the matter-energy-momentum tensor, leading to the concept of \( f(Q,T) \) gravity ~\cite{Xu:2019sbp}, widely studied in the literature~\cite{Tayde:2024,Pati:2022}. The first cosmological solutions within the $f(Q)$ framework were presented in~\cite{BeltranJimenez:2019tme}. Afterwards, a number of studies explored different cosmological and physical implications of the theory. The authors in~\cite{Mandal:2020lyq} examined the energy conditions for various $f(Q)$ models and discussed their consistency with late-time cosmic acceleration, whereas cosmographic properties were analysed in~\cite{Mandal:2020buf}. The behavior of matter perturbations, including the growth index, has been investigated in~\cite{Khyllep:2021pcu}. Additional aspects such as geodesic deviation~\cite{Beh:2021wva}, matter coupling~\cite{Harko:2018gxr}, quantum cosmology~\cite{Dimakis:2021gby}, and holographic DE~\cite{Shekh:2021ule} have been discussed, and several further developments related to this framework can be found in~\cite{De:2022shr, Lazkoz:2019sjl, Frusciante:2021sio, Barros:2020bgg, Zhao:2021zab}.  

On the theoretical side, Einstein originally attributed gravity to the torsion of spacetime in metric teleparallel theories~\cite{unzicker2005translation}. In teleparallel frameworks, one may build either the torsion scalar $T$ or the non-metricity scalar $Q$, and obtain an equivalent description of Einstein gravity by replacing the Ricci scalar in the Einstein--Hilbert action with $T$ or $Q$, respectively. These theories remain dynamically equivalent to GR up to a boundary term, and consequently still require dark components to explain cosmic acceleration. Motivated by this issue, the $f(T)$ and $f(Q)$ extensions were introduced, offering modified field equations that remain second order, unlike the fourth-order field equations appearing in $f(R)$ gravity~\cite{Linder:2010py}. Moreover, certain conceptual ambiguities in $f(T)$ formulations~\cite{Li:2010cg, Golovnev:2021htv, Krssak:2018ywd, Golovnev:2021lki} are absent in the $f(Q)$ approach, and detailed comparisons between the two may be found in~\cite{BeltranJimenez:2019esp, DAmbrosio:2021pnd, Capozziello:2022zzh, Beh:2021wva, Lu:2021wif, De:2022shr}. Interestingly, the work done in~\cite{Anagnostopoulos:2021ydo} showed that $f(Q)$ gravity can challenge the standard $\Lambda$CDM cosmology. In particular, even with the same number of free parameters as $\Lambda$CDM, the theory may not admit $\Lambda$CDM as its limiting case, and thus could offer a possible route to alleviating the cosmological constant problem. 

Most existing studies focus on simple extensions such as linear deviations, power-law forms, exponential forms, or logarithmic corrections. While these models provide useful insights, they often lack the flexibility to simultaneously capture non-trivial deviations from $\Lambda$CDM at both background and perturbative levels. Moreover, many analysis restrict the parameter space or impose strong priors, potentially overlooking viable regions consistent with current high-precision datasets. Motivated by these gaps, we consider in this work a new novel functional form of $f(Q)$ introduced by Starobinsky~\cite{Starobinsky:2007}. This form is explored in extensions of GR and teleparallel gravity formalism~\cite{Yang:2011}, and this work is analysing the model in modified symmetric-teleparallel $f(Q)$ gravity formalism. In this work, we demonstrate the observational constraints on the model parameters from joint analysis using the recent datasets, including cosmic chronometers (CC), PantheonPlus SHOES (PPS) sample, and the DE Spectroscopic Instrument (DESI) BAO Data Release 2 (DR2). The purpose is to examine whether such a functional form can offer an improved description of the Universe's expansion history in the modified symmetric teleparallel gravity formalism. The rest of this paper is structured as follows. In Section~\ref {gravity_formalism}, we outline the theoretical framework of symmetric teleparallel gravity. Section~\ref {data} describes the observational datasets and the Monte Carlo Markov Chain (MCMC) methodology employed. In Section~\ref {model_analysis}, we present the analysis of the model using different cosmological parameters. Finally, Section~\ref {conclusion} summarizes our findings and outlines future research directions.

\section{$f(Q)$ gravity Formalism}\label{gravity_formalism}
In this section, we examine the comprehensive formulation of symmetric teleparallelism, particularly focusing on its extended version known as the modified \(f(Q)\) theory. We start with a four-dimensional Lorentzian manifold \(\mathcal{M}^4\), defined by the metric tensor \(g_{\mu\nu}\) in a specific coordinate system \(\{x^0,x^1,x^2,x^3\}\), along with a (generally non-tensorial) affine connection \(\Gamma^\alpha{}_{\mu\nu}\), which dictates the covariant derivative \(\nabla\) and addresses the three main aspects of the spacetime geometry related to this connection: curvature, torsion, and non-metricity. However, if we limit our consideration to cases where both the torsion tensor \(T^\alpha{}_{\mu\nu}\) and the non-metricity tensor \(Q_{\alpha\mu\nu}\) vanish with respect to the connection, we can conclude that there exists a unique connection, the Levi--Civita connection, which is completely determined by the metric \(g_{\mu\nu}\). Explicitly, the Levi--Civita connection coefficients (Christoffel symbols),  contorsion tensor,
 and disformation tensor are given by,
\begin{align}
\Gamma^{\sigma}{}_{\mu\nu}
    &= \frac{1}{2} g^{\sigma\lambda}
      \left( \partial_{\mu} g_{\lambda\nu}
           + \partial_{\nu} g_{\lambda\mu}
           - \partial_{\lambda} g_{\mu\nu} \right), \\[6pt]
K^{\sigma}{}_{\mu\nu}
    &= \frac{1}{2} T^{\sigma}{}_{\mu\nu}
      + T^{\ \ \sigma}_{(\mu\ \ \nu)}, \\[6pt]
L^{\sigma}{}_{\mu\nu}
    &= -\frac{1}{2} g^{\sigma\lambda}
      \left( Q_{\mu\lambda\nu}
           + Q_{\nu\lambda\mu}
           - Q_{\lambda\mu\nu} \right).
\end{align}
where, to explore the cosmological aspects of nonmetricity gravity, let us consider the most comprehensive representation of the affine connections can be considered as,
\begin{equation}
\hat{\Gamma}^\sigma_{\ \mu\nu}
= \Gamma^\sigma_{\ \mu\nu}
+ K^\sigma_{\ \mu\nu}
+ L^\sigma_{\ \mu\nu}.
\end{equation}
The non-metricity tensor \( Q_{\sigma\mu\nu} \) is defined as
\begin{align}
Q_{\sigma\mu\nu} &= \nabla_{\sigma} g_{\mu\nu},
\end{align}
and its associated traces are given by,
\begin{align}
Q_{\sigma} &= Q^{\mu}{}_{\sigma\mu}, \\
\tilde{Q}_{\sigma} &= Q^{\mu}{}_{\sigma\mu}.
\end{align}
These are used to obtain the superpotential tensor as,
\begin{equation}
P^\sigma_{\ \mu\nu}
= -\frac{1}{4} Q^\sigma_{\ \mu\nu}
+ \frac{1}{2} Q_{(\mu\ \nu)}^{\ \ \ \sigma}
- \frac{1}{4}
\left( Q^\sigma + \tilde{Q}^\sigma \right) g_{\mu\nu}
- \frac{1}{4} \delta^\sigma_{(\mu} Q_{\nu)}.
\end{equation}
The nonmetricity trace can be calculated using the formula,
\begin{equation}
Q = - Q_{\sigma\mu\nu}
P^{\sigma\mu\nu}.
\end{equation}
The action for the $f(Q)$ gravity can be presented as~\cite{Jimenez:2017tkx}, 
\begin{equation}
S = -\frac{1}{2\kappa^{2}}
\int \left[\, Q + f(Q) + \mathcal{L}_{m} \,\right] \sqrt{-g}\, d^{4}x .\label{aceq}
\end{equation}
here, $\mathcal{L}_{m}$ denotes the matter Lagrangian, $Q$ is the nonmetricity scalar, $g$ is the determinant of the metric tensor.
The variation of the action formula mentioned in Eq. \ref{aceq} with respect to the metric, the field
equation for modified nonmetricity gravity $f(Q)$ is obtained as,
\begin{widetext}
\begin{equation}
\frac{2}{\sqrt{-g}} \nabla_{\sigma} \!\left[(1+f_{Q})\sqrt{-g}\, 
P^{\sigma}{}_{\mu\nu} \right]
+ \frac{1}{2}\left(Q + f(Q)\right) g_{\mu\nu}
+ (1+f_{Q})\left(
P_{\mu\sigma\lambda} Q_{\nu}{}^{\sigma\lambda}
- 2 Q_{\sigma\lambda\mu} P_{\nu}{}^{\sigma\lambda}
\right)
= T_{\mu\nu}.
\end{equation}
\end{widetext}
where the energy--momentum tensor for matter is now
defined as
$T_{\mu\nu} \equiv - \frac{2}{\sqrt{-g}}
\frac{\delta(\sqrt{-g}\,\mathcal{L}_m)}{\delta g^{\mu\nu}}$ and $f_Q = \dfrac{df}{dQ}$. 
To analyse $f(Q)$ gravity in a cosmological context, we adopt the spatially
flat FLRW spacetime, characterized by a specific metric.
\begin{equation}
ds^2 = -dt^2 + a(t)^2\, \delta_{ij}\, dx^i dx^j, \quad (i,j=1,2,3)
\end{equation}
The scale factor $a(t)$ describes the expansion of the universe. The Hubble parameter, which quantifies the rate of expansion, is expressed as $H = \frac{\dot{a}}{a}$. The non-metricity scalar $Q$ serves a function similar to that of the Ricci scalar $R$ in GR. For the above metric, the non-metricity scalar can be simplified to $Q = 6H^2$. The field equations that describe the dynamics of the Universe in $f(Q)$ gravity are expressed as follows,
\begin{eqnarray}
3 H^{2}
&=& \frac{f}{2} 
- Q f_Q
+ \rho_{m} + \rho_{r} , \label{FE1}
\\[8pt]
2 \dot{H} + 3 H^{2}
&=& \frac{f}{2} 
- 2 \dot{H} f_Q
- Q f_Q \nonumber\\
&-& 2 H \dot{Q} f_{QQ}
- \frac{ \rho_{r}}{3}. \label{FE2}
\end{eqnarray}
The general Friedmann equations are as noted below,
\begin{eqnarray}
3H^{2}
&=& \rho_{m} + \rho_{r} + \rho_{\text{DE}},
\\[8pt]
2\dot{H} + 3H^{2}
&=& - \frac{ \rho_{r}}{3} - p_{\text{DE}}.
\end{eqnarray}
On comparing with Eqs. \ref{FE1} and \ref{FE2}, we can acquire the contribution of density, pressure, and the Equation of State (EoS) parameter for DE as follows,

\begin{eqnarray}
\rho_{\text{DE}}
&=& \frac{f}{2}  - Q f_Q ,
\\[8pt]
p_{\text{DE}}
&=& 2 H \dot{Q} f_{QQ}
+ 2\dot{H} f_{Q}-\frac{f}{2}+Qf_{Q},
\\[8pt]
w_{\text{DE}}
&=& \frac{p_{\text{DE}}}{\rho_{\text{DE}}}
= -1
+ \frac{4 \dot{H}\left(f_{Q} +2 Q f_{QQ}\right)}
{f -2 Q f_Q }.
\end{eqnarray}
It can be easily verified that the contribution from matter and radiation obeys the conservation equations as stated below,
\begin{align}
\dot{\rho}_m + 3H\,\rho_m &= 0, \label{eq:rho_m}\\[6pt]
\dot{\rho}_r + 4H\,\rho_r &= 0 \label{eq:rho_r}
\end{align}
For the DE contribution, it can be stated as follows, 
\begin{align}
\dot{\rho}_{DE} + 3H\,\left(\rho_{DE}+p_{DE}\right) &= 0 \label{eq:rho_DE}
\end{align}
\subsection{The $H(z)$ Parametrization}
The rate of expansion of the universe is commonly characterized by parameterizing the deceleration parameter $q(z)$, which can take the form,
\begin{equation}\label{eq:q_param}
q(z) = \tau_0 + \tau_1 \mu(z), 
\end{equation}
where $\tau_0$ and $\tau_1$ are constants, and $\mu(z)$ is a function of the redshift $z$. Various forms for $\mu(z)$ have been suggested in the literature, each potentially solving different cosmological issues. However, certain parameterizations do not accurately project the universe’s future evolution, while some are limited to small redshifts ($z \ll 1$). Gong and Wang~\cite{Gong:2006} introduced a particular parameterization of the deceleration parameter as a function of redshift, expressed as,
\begin{equation}\label{eq:q_gong_wang}
q(z) = \frac{mz + n}{(1 + z)^2}+\frac{1}{2}. 
\end{equation}
They showed that this parameterization offers a superior fit to observational data compared to the $\Lambda$CDM model in certain ranges of redshift (for instance, around $z\approx 0.2$). By utilizing Eq. \ref{eq:q_gong_wang} and the definition of the deceleration parameter $q(z)=-1+(1+z)\frac{1}{H(z)}\frac{dH(z)}{dz}$, we can formulate an expression for the Hubble parameter as a function of the redshift $z$~\cite{Dubey:2025} as follows,
\begin{equation}\label{eq:H_of_z}
H(z) = \exp\!\left[\frac{n}{2} + \frac{mz^2 - n}{2(1 + z)^2}\right] H_0 (1 + z)^{\frac{3}{2}} .
\end{equation}
In the following sections, we present observational datasets used for the model parameter estimation, followed by an analysis of cosmological parameters like the behavior of density, EoS parameter, and deceleration parameter for the constrained values of the model parameters. 

\section{Dataset Description and Parametric Constraints} \label{data}
We summarize here the observational datasets used in our analysis and the parametric constraints inferred from them. We briefly describe the datasets and then outline the MCMC framework adopted to construct the likelihood and explore the parameter space, leading to robust posterior constraints on the model parameters.
\subsection{Cosmic Chronometer (CC)}
The expansion history of the Universe can be directly probed through the Hubble parameter $H(z)$, and one of the most reliable techniques to obtain model-independent measurements is the \textit{Cosmic Chronometer} (CC) approach. This technique estimates $H(z)$ by determining the differential age evolution of passively evolving, early-type galaxies~\cite{Simon:2004tf,Stern:2009ep,Moresco:2015cya,Ratsimbazafy:2017vga}. Since these galaxies evolve slowly after their star-formation epoch, their age difference directly traces the derivative $dz/dt$, allowing $H(z)$ to be measured through  
\begin{equation}
    H(z) = -\frac{1}{1+z}\,\frac{dz}{dt}.
\end{equation}
Here, we employ $31$ CC based $H(z)$ measurements spanning the redshift interval $0.07 \leq z \leq 1.965$. These data offer valuable constraints on the cosmic expansion rate at low and intermediate redshifts. The complete list of measurements, along with their sources, is presented in Table~\ref{OHDdata31}.
\begin{table*}[ht]
\centering
\caption{The $31$ $H(z)$ measurements from the CC method used in this study in units of $\mathrm{Km\,s^{-1} Mpc^{-1}}$.}
\label{OHDdata31}

\renewcommand{\arraystretch}{1.5}
\setlength{\tabcolsep}{9pt}

\begin{tabular}{l c c c c @{\hspace{1cm}} l c c c c}
\hline\hline
S.No & $z$ & $H(z)$ & Error & Reference &
S.No & $z$ & $H(z)$ & Error & Reference \\
\hline
1  & 0.07    & 69.0  & 19.6 & \cite{zhang2016test} &
17 & 0.4783 & 80.9  & 9.0  & \cite{moresco20166} \\
2  & 0.09    & 69.0  & 12.0 & \cite{Simon:2004tf} &
18 & 0.48   & 97.0  & 62.0 & \cite{Stern:2009ep} \\
3  & 0.12    & 68.6  & 26.2 & \cite{zhang2016test} &
19 & 0.593  & 104.0 & 13.0 & \cite{Moresco:2012jh} \\
4  & 0.17    & 83.0  & 8.0  & \cite{Simon:2004tf} &
20 & 0.68   & 92.0  & 8.0  & \cite{Moresco:2012jh} \\
5  & 0.179   & 75.0  & 4.0  & \cite{Moresco:2012jh} &
21 & 0.781  & 105.0 & 12.0 & \cite{Moresco:2012jh} \\
6  & 0.199   & 75.0  & 5.0  & \cite{Moresco:2012jh} &
22 & 0.875  & 125.0 & 17.0 & \cite{Moresco:2012jh} \\
7  & 0.20    & 72.9  & 29.6 & \cite{zhang2016test} &
23 & 0.88   & 90.0  & 40.0 & \cite{Stern:2009ep} \\
8  & 0.27    & 77.0  & 14.0 & \cite{Simon:2004tf} &
24 & 0.9    & 117.0 & 23.0 & \cite{Simon:2004tf} \\
9  & 0.28    & 88.8  & 36.6 & \cite{zhang2016test} &
25 & 1.037  & 154.0 & 20.0 & \cite{Moresco:2012jh} \\
10 & 0.352   & 83.0  & 14.0 & \cite{Moresco:2012jh} &
26 & 1.3    & 168.0 & 17.0 & \cite{Simon:2004tf} \\
11 & 0.3802  & 83.0  & 13.5 & \cite{moresco20166} &
27 & 1.363  & 160.0 & 33.6 & \cite{Moresco:2015cya} \\
12 & 0.40    & 95.0  & 17.0 & \cite{Simon:2004tf} &
28 & 1.43   & 177.0 & 18.0 & \cite{Simon:2004tf} \\
13 & 0.4004  & 77.0  & 10.2 & \cite{moresco20166} &
29 & 1.53   & 140.0 & 14.0 & \cite{Simon:2004tf} \\
14 & 0.4247  & 87.1  & 11.2 & \cite{moresco20166} &
30 & 1.75   & 202.0 & 40.0 & \cite{Simon:2004tf} \\
15 & 0.4497  & 92.8  & 12.9 & \cite{moresco20166} &
31 & 1.965  & 186.5 & 50.4 & \cite{Moresco:2015cya} \\
16 & 0.47    & 89.0  & 50.0 & \cite{Ratsimbazafy:2017vga} & & & & & \\
\hline\hline
\end{tabular}
\end{table*}

The likelihood function for the CC is defined as
\begin{align}\label{chi^2}
\chi^2_{\rm CC}(z) = \sum_{i=1}^{31} \left[\frac{H_{\text{t}}\left(\alpha, H_0, z_i\right) - H_{\text{ob}}\left(z_i\right)}{\sigma_H\left(z_i\right)}\right]^2,
\end{align}
where $H_{\text{t}}\left(\alpha, H_0, z_i\right)$ is the theoretical value determined from our considered model at different redshifts $z_i$, and $H_{\text{ob}}\left(z_i\right)$ corresponds to the observed Hubble parameter, while $\sigma_H$ indicates the measurement error.

\subsection{PantheonPlus SH0ES}
The PantheonPlus SH0ES compilation comprises standardized Type Ia supernovae (SNe~Ia), which provide a reliable probe of the late-time expansion due to their precisely calibrated intrinsic luminosities. This sample includes $1701$ light curves corresponding to $1550$ different SNe~Ia events distributed across the redshift range from $z=0.001$ to $z=2.26$. In this dataset, the primary observable is the apparent magnitude $m_B$, whose theoretical value at redshift $z$ is calculated as  
\begin{equation}
m_B(z) = 5 \log_{10}\left[ \frac{d_L(z)}{1~\mathrm{Mpc}} \right] + 25 + M_B .
\end{equation}
The apparent magnitude depends on the predicted luminosity distance $d_L(z)$ and on the absolute magnitude $M_B$, which characterizes the intrinsic brightness of SNe~Ia after light-curve standardization. Once calibrated, $M_B$ is treated as a redshift-independent quantity, enabling all supernovae to be compared on the same physical scale.
The luminosity distance itself carries the imprint of the expansion history and is defined as  
\begin{equation}
d_L(z) = (1+z)\, \int_0^{z} \frac{c\, dz'}{H(z')} .
\end{equation}
It increases with redshift according to the underlying $H(z)$, and therefore connects supernova brightness directly to the cosmic expansion rate. The observed supernova brightness is usually expressed through the distance modulus,
\begin{equation}
\mu(z) \equiv m_B(z) - M_B = 5\log_{10}\!\left[ \frac{d_L(z)}{1~\mathrm{Mpc}} \right] + 25 ,
\end{equation}
which is the quantity supplied by the PantheonPlus SH0ES catalogue along with its full covariance matrix.

We incorporate the distance modulus measurements of Type Ia supernovae from the PantheonPlus sample~\cite{brout2022pantheon+}, together with the SH0ES Cepheid host galaxy distance calibrations~\cite{brout2022pantheon+}. These measurements have been considered from $18$ different surveys such as CfA1-CfA4~\cite{Riess:1998dv, Jha:2005jg, Hicken:2009df, Hicken:2012zr}, CSP~\cite{krisciunas2017carnegie}, SOUSA~\cite{brown2014sousa}, CNIa0.02~\cite{Chen:2020qnp}, Foundation~\cite{foley2018foundation}, and LOSS~\cite{stahl2019lick} which are focusing in the range of $0.01$ to $0.1$. DES~\cite{brout2019first}, SNLS~\cite{SDSS:2014iwm}, SDSS~\cite{Sako:2011um}, and PS1~\cite{Pan-STARRS1:2017jku} focus on $z$ greater than $0.1$ and SCP, GOODS, HDFN, CANDLES/CLASH released $z$ greater than $1.0$ data~\cite{riess2004type, Riess:2006fw, SupernovaCosmologyProject:2011ycw}. The corresponding likelihood is constructed as  
\begin{equation}
\label{chi2_PPS}
\chi^2_{\rm PPS}(z)=\sum_{j,k=1}^{1701}\, \Delta\mu_j \left(C_{SN}^{-1}\right)_{j,k} \Delta\mu_k ,
\end{equation}
where $\Delta\mu = \mu_{\rm th} - \mu_{\rm obs}$ denotes the residual between the model-predicted and observed distance moduli. The matrix $C_{SN}$ represents the full covariance of the \texttt{PPS} sample, incorporating both statistical and systematic uncertainties.

\subsection{DESI BAO}
Baryon Acoustic Oscillations arise from acoustic wave propagation in the photon--baryon plasma of the early Universe, leaving a fixed comoving imprint on the distribution of galaxies. This produces a characteristic clustering scale that functions as a standard ruler for cosmological distances.
The BAO standard ruler is the sound horizon at the drag epoch, $r_d$, which corresponds to the maximum distance travelled by acoustic waves before baryon--photon decoupling. It is computed from the sound speed of the coupled plasma as
\begin{equation}
r_d = \int_{z_d}^{\infty} \frac{c_s(z)}{H(z)} \, dz .
\end{equation}
In this work, we utilize BAO measurements from DESI DR2~\cite{karim2025desi}, which provide the spectroscopic mapping of large-scale structure. DESI measured redshifts for over $30$ million galaxies and quasars within its first three years, of which $\sim 14$ million high-fidelity redshifts are used for BAO clustering analysis. The survey operates in bright and dark modes, targeting BGS galaxies at low redshift and LRGs, ELGs, and QSOs at higher redshifts, thereby providing BAO coverage over $0.1 < z < 2.330$. The DR2 footprint includes $6671$ dark and $5171$ bright tiles, improving sky coverage by factors of $2.4$ and $2.3$ relative to DR1~\cite{adame2024desi} and substantially increasing completeness across all tracers. After reconstruction and two-point clustering estimation, DESI constrains the cosmological distance ratios $D_M/r_d$ and $D_H/r_d$, enabling precise measurements of the expansion history. However, BAO measurements alone constrain only the combined quantity $H_0\, r_d$, and external early-Universe priors are required to separately determine $H_0$ and $r_d$. The considered data points are mentioned in Table~\ref{DESI BAO}.
\begin{table*}[ht]
\centering
\caption{The $9$ points from DESI BAO DR2 measurements used in the present analysis.}
\label{DESI BAO}

\renewcommand{\arraystretch}{1.5}
\setlength{\tabcolsep}{12pt}

\begin{tabular}{l c c c c}
\hline\hline
Tracer & $z_{\rm eff}$ & $D_M/r_d$ & $D_H/r_d$ & $D_V/r_d$ \\
\hline
{\scriptsize BGS}  & $0.295$ & $-$ & $-$ & $7.942 \pm 0.075$ \\
{\scriptsize LRG1} & $0.510$ & $13.588 \pm 0.167$ & $21.863 \pm 0.425$ & $12.720 \pm 0.099$ \\
{\scriptsize LRG2} & $0.706$ & $17.351 \pm 0.177$ & $19.455 \pm 0.330$ & $16.050 \pm 0.110$ \\
{\scriptsize LRG3} & $0.922$ & $21.648 \pm 0.178$ & $17.577 \pm 0.213$ & $19.656 \pm 0.105$ \\
{\scriptsize ELG1} & $0.955$ & $21.707 \pm 0.335$ & $17.803 \pm 0.297$ & $20.008 \pm 0.183$ \\
{\scriptsize LRG3+ELG1} & $0.934$ & $21.576 \pm 0.152$ & $17.641 \pm 0.193$ & $19.721 \pm 0.091$ \\
{\scriptsize ELG2} & $1.321$ & $27.601 \pm 0.318$ & $14.176 \pm 0.221$ & $24.252 \pm 0.174$ \\
{\scriptsize QSO} & $1.484$ & $30.512 \pm 0.760$ & $12.817 \pm 0.516$ & $26.055 \pm 0.398$ \\
{\scriptsize Ly-$\alpha$ QSO} & $2.330$ & $38.988 \pm 0.531$ & $8.632 \pm 0.101$ & $31.267 \pm 0.256$ \\
\hline\hline
\end{tabular}
\end{table*}

\subsection{Parametric Constraints}
In this work, we perform a joint cosmological analysis using the DESI BAO DR2 measurements in combination with CC and PPS datasets. For sampling the cosmological parameter space, we use the \texttt{emcee} Python package~\cite{foreman2013emcee}, which implements an affine-invariant ensemble MCMC sampler. This algorithm evolves an ensemble of walkers in parameter space, updating each walker by proposing moves that are informed by the positions of the other walkers. After an initial burn-in phase, the retained chain samples are used to estimate the posterior distributions and credible intervals for the model parameters. 
In our analysis, we adopt uniform (flat) priors on the free parameters of the model, chosen to be sufficiently wide so as not to bias the parameter estimation. Specifically, the prior ranges are given by: $H_0 \in [40,100]$,   $m \in [-5,5]$, and $ n \in [-5,5]$.
Finally, the resulting MCMC samples are post-processed using the \texttt{GetDist} Python package to compute marginalized constraints and to generate confidence contours for the cosmological parameters.

\begin{figure*}[ht]
    \centering
    \includegraphics[width=0.8\textwidth]{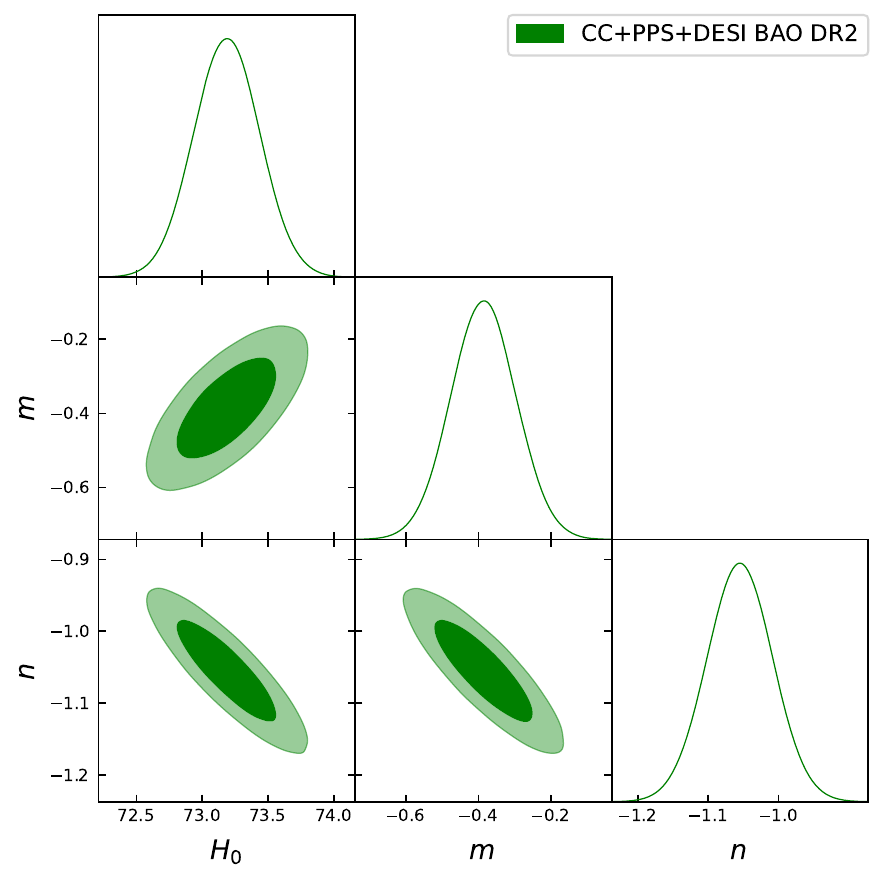}
    \caption{One-dimensional marginalized posterior distributions and
two-dimensional joint contours of the model parameters.}
    \label{joint}
\end{figure*}

We constrain the model parameters using CC + PPS + DESI BAO DR2 and summarize the results in Table~\ref{params}. The best fit value of Hubble constant is obtained as $73.19\pm 0.25~~\mathrm{Km\,s^{-1} Mpc^{-1}}$ at 68\% CL which is consistent with SH0ES measurement $H_0 =73.04 \pm 1.04~~\mathrm{Km\,s^{-1} Mpc^{-1}}$~\cite{riess2004type}. The present deceleration parameter $q_0$ depends on the parameter $n$ with the relation $q_0=\frac{1}{2}+n$, the conditions on $(m,n)$ define the expansion behavior of the universe. If $m>0$ and $n>0$, the model predicts a decelerating phase. For $m<0$ and $n>-1/2$, the universe remains in deceleration at the present epoch but would have undergone acceleration in the past. For $m<0$ and $n<-1/2$, the model predicts the accelerating phase at the present epoch. In our analysis, we obtain the model parameter values as $m = -0.386\pm 0.090$ and $n = -1.055\pm 0.047$ at $68\%$ CL, which clearly indicates the universe is in an accelerating phase. One can see the likelihood contour plots for our model in the Fig. \ref{joint}.

\begin{table*}[ht]
\centering
\caption{Constraints on the parameters $H_0$, $m$, and $n$ obtained from the combined dataset: CC+PPS+DESI BAO DR2.}
\label{params}

\renewcommand{\arraystretch}{1.6}
\setlength{\tabcolsep}{10pt}

\begin{tabular}{l c c c}
\hline\hline
\textbf{Parameter} 
& \textbf{68\% C.L.} 
& \textbf{95\% C.L.} 
& \textbf{99\% C.L.} \\
\hline
$H_0~(\mathrm{km\,s^{-1}\,Mpc^{-1}})$ 
& $73.19 \pm 0.25$ 
& $73.19 \pm 0.49$ 
& $73.19 \pm 0.65$ \\

$m$ 
& $-0.386 \pm 0.090$ 
& $-0.39 \pm 0.18$ 
& $-0.39 \pm 0.23$ \\

$n$ 
& $-1.055 \pm 0.047$ 
& $-1.055 \pm 0.092$ 
& $-1.06 \pm 0.12$ \\
\hline\hline
\end{tabular}
\end{table*}

\section{Model Analysis Using Cosmological Parameters }\label{model_analysis}
\begin{figure}[ht]
    \centering
    \includegraphics[width=0.45\textwidth]{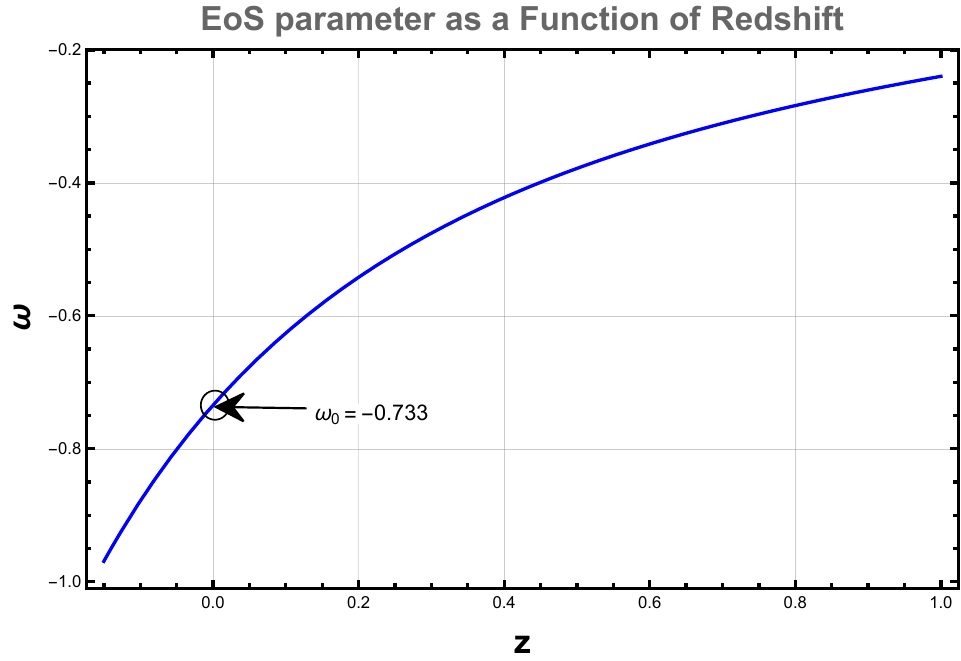}
    \caption{Evolution of the EoS parameter $\omega$ as function of redshift $z$, for the constrained coefficients from Fig. \ref{joint}.}
    \label{fig_eos}
\end{figure}

\begin{figure}[H]
    \centering
    \includegraphics[width=0.45\textwidth]{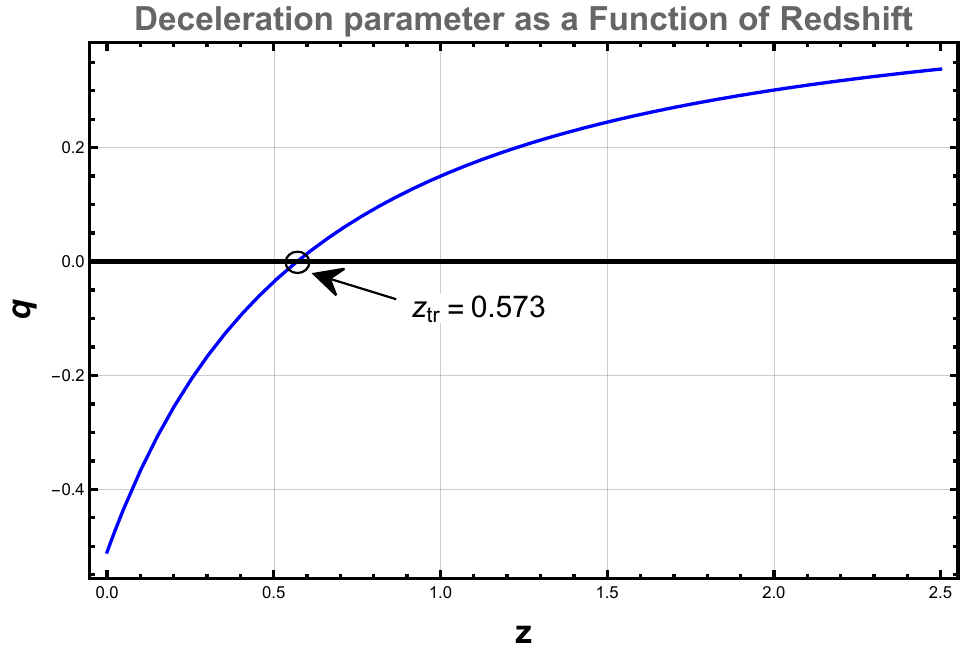} \hfill
    \caption{Evolution of the deceleration parameter $q$ as functions of redshift $z$ for the constrained coefficients from Fig. \ref{joint}.}
    \label{fig:q}
\end{figure}

The Starobinsky model that we take into consideration in this study is of the form 
\begin{align}
    f(\text{Q})=Q-\alpha  Q_0 \left[\left(\frac{Q^2}{Q_{0}^2}+1\right)^{-\beta }-1\right]
\end{align}
where $\alpha$, $\beta$ are the positive constants, $Q_0$ is the order of the present Hubble parameter. The same form is examined in the framework of $f(R)$ gravity theory~\cite{Starobinsky:2007}, and has also been thoroughly tested in the context of $f(T)$ gravity~\cite{Yang:2011}. To analyse different cosmological parameters in the $f(Q)$ gravity formalism adopted here, we have presented the energy density, EoS parameter, and the deceleration parameter profile in terms of the redshift in Figs. \ref{fig_eos} and \ref{fig:q}. 

The EoS parameter $\omega$ characterizes the nature of the expansion of the universe. The three possibilities for acceleration of the universe are the cosmological constant for which $\omega=-1$, the phantom regime for which $\omega < -1$, and the quintessence regime where $-1<\omega<-1/3$. Fig. \ref{fig_eos} indicates $\omega$ remains above $-1$ throughout the redshift range and approaches $\omega  \approx -1$ in the far future. 
\begin{figure}[ht]
    \centering
    \includegraphics[width=0.45\textwidth]{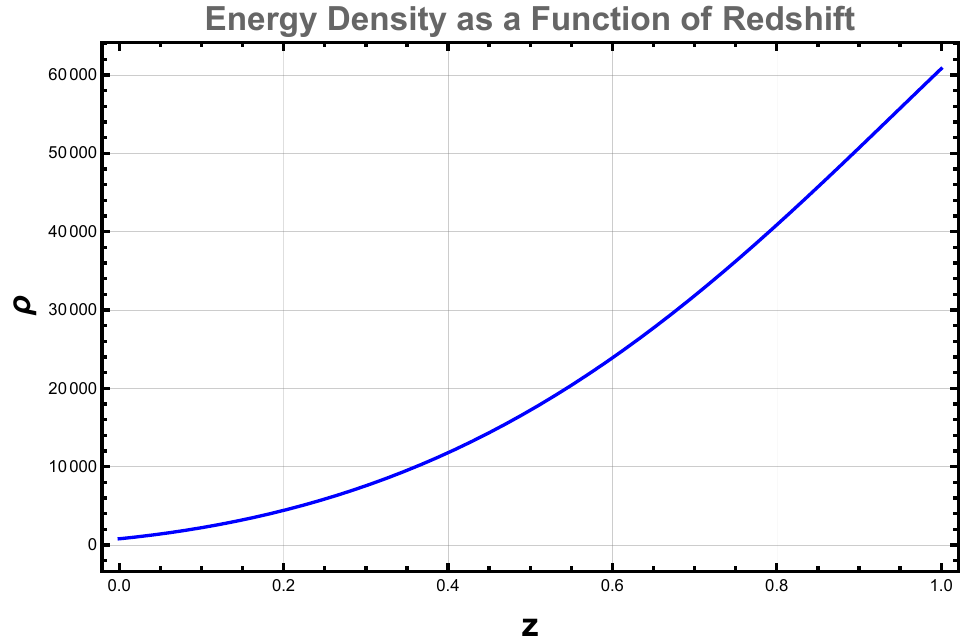} \hfill
    \caption{Evolution of the energy density $\rho$ as functions of redshift $z$, for the constrained coefficients from Fig. \ref{joint}.}
    \label{fig_density}
\end{figure}

 This behavior indicates the DE in our model exhibits a quintessence regime with $\omega_{0} = -0.73 \pm 0.3$, which agrees with the observational studies~\cite{Capozziello:2014} at present. The deceleration parameter describes the expansion of the universe. A positive value of $q$ signifies the decelerated expansion as in the matter-dominated era, where a negative $q$ indicates accelerated expansion of the universe as in the DE-dominated era.  From Fig. \ref{fig:q}, the current deceleration parameter yields $\approx -0.51 \pm 0.6$. This confirms the ongoing accelerated expansion of the universe, with the negative $q_0$ indicating the present dominance of DE. Moreover, the transition occurs at a redshift, $z_{tr} \approx 0.573$, which is consistent with the widely accepted value of the transition redshift inferred from Planck-CMB 2018 observations within the $\Lambda$CDM model~\cite{Planck:2018vyg}. From recent literature investigations carried out for different objectives~\cite{Maurya:2024nxx, Zhadyranova:2024lyz, Myrzakulov:2024esv, Yang:2019fjt}, we have noted that transition redshift $z_{tr}$ lies in the range $z_{tr} \approx 0.50-0.70$, and our observed $z_{tr}$ value lies in the same range, indicating that our finding is consistent with the existing studies.

\subsection{Energy Conditions}\label{ec}
\begin{figure}[H]
    \centering
\includegraphics[width=0.45\textwidth]{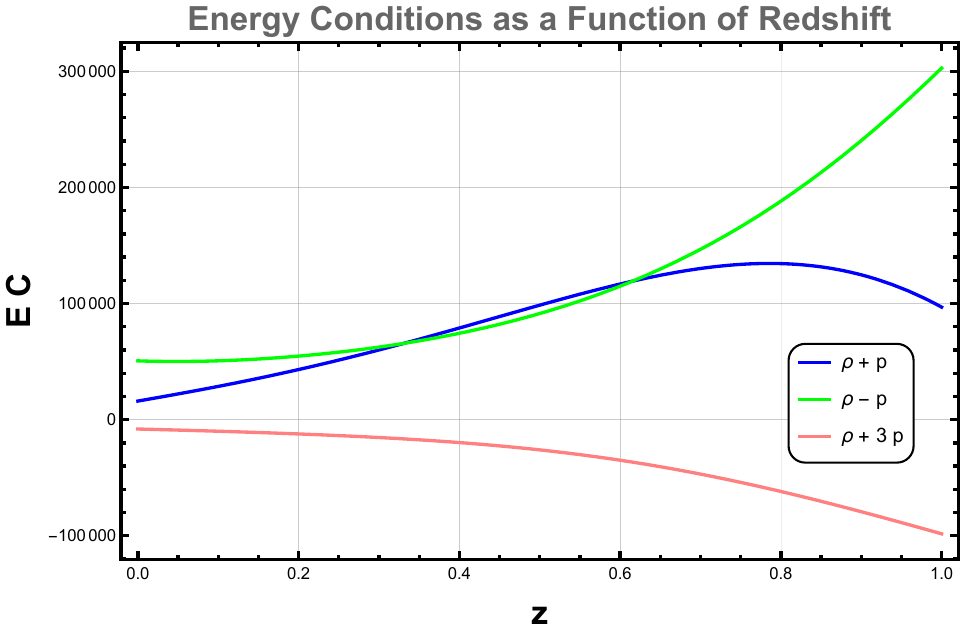}
    \caption{Evolution of energy conditions: Null [NEC, $\rho+p$], Dominant [DEC. $\rho-p$], Strong [SEC, $\rho+3p$] as functions of redshift $z$, for the constrained coefficients from Fig. \ref{joint}.}
    \label{fig:energy-conditions}
\end{figure}

Energy conditions establish constraints on the energy-momentum tensor that are independent of the coordinate system. The primary conditions are~\cite{Raychaudhuri-1955, Carroll-2019, Curiel2017}:  
\begin{itemize}
    \item \textbf{Weak Energy Condition (WEC):}  
    $T_{ij}t^{i}t^{j}\geq 0$ for any timelike vector $t^{i}$.  
    For a perfect fluid:  
    \begin{equation}
        T_{ij}u^{i}u^{j}=\rho, \quad 
        T_{ij}\xi^{i}\xi^{j}=(\rho+p)(u_{i}\xi^{i})^{2}.
    \end{equation}
    This implies that $\rho \geq 0$ and $\rho+p \geq 0$.  

    \item \textbf{Null Energy Condition (NEC):}  
    $T_{ij}\xi^{i}\xi^{j}\geq 0$ for any null vector $\xi^{i}$.  
    This condition is equivalent to $\rho+p \geq 0$.  

    \item \textbf{Strong Energy Condition (SEC):}  
    $T_{ij}t^{i}t^{j}-\tfrac{1}{2}T^{k}_{~k}t^{l}t_{l}\geq 0$.  
    This is equivalent to $\rho+p \geq 0$ and $\rho+3p \geq 0$.  
    It implies that gravity is attractive.  

    \item \textbf{Dominant Energy Condition (DEC):}  
    $T_{ij}t^{i}t^{j}\geq 0$ and $T^{ij}t_{i}$ is non-spacelike.  
    For a perfect fluid: $\rho \geq |p|$.  
\end{itemize}
    The energy conditions essentially serve as the boundary conditions that influence the evolution of the cosmos and are presented in Fig. \ref{fig:energy-conditions}. Additionally, due to the fundamental structure of spacetime, the energy conditions characterise gravitational attraction. The NEC remains positive throughout cosmic evolution, signifying that the energy density is non-negative. The DEC is consistently upheld, ensuring that the energy density surpasses the pressure and that energy propagation remains within causal limits. The SEC, which is breached in the early and later epochs. This breach aligns with the observed accelerated expansion and implies a divergence from traditional matter-dominated models. Analyzing these conditions enables the determination of the characteristics of matter and energy in the Universe, which is crucial for comprehending its accelerated expansion and the influence of DE.
\subsection{$Om (z)$ Diagnostic}\label{omz}
The $Om(z)$ diagnostic has been established as an alternative method to examine the accelerated expansion of the Universe based on the phenomenological assumption that the EoS is $p = \rho \omega$, treating the universe as a perfect fluid. The $Om(z)$ diagnostic offers a way to test the $\Lambda$CDM model without assumptions~\cite{Om_Sahni_2008}. Additionally, literature provides evidence of its sensitivity to the EoS parameter~\cite{Ding_2015, Zheng_2016,Qi_2018}. The slope of the $Om(z)$ function varies among DE models; a positive slope signifies the phantom phase where $\omega < -1$, while a negative slope indicates the quintessence region where $\omega > -1$. The definition of the $Om(z)$ diagnostic can be expressed as follows:
\begin{equation}
\Omega_m(z) = \frac{E^2(z) - 1}{(1+z)^3 - 1}
\end{equation}
\begin{figure}[H]
    \centering
    \includegraphics[width=0.45\textwidth]{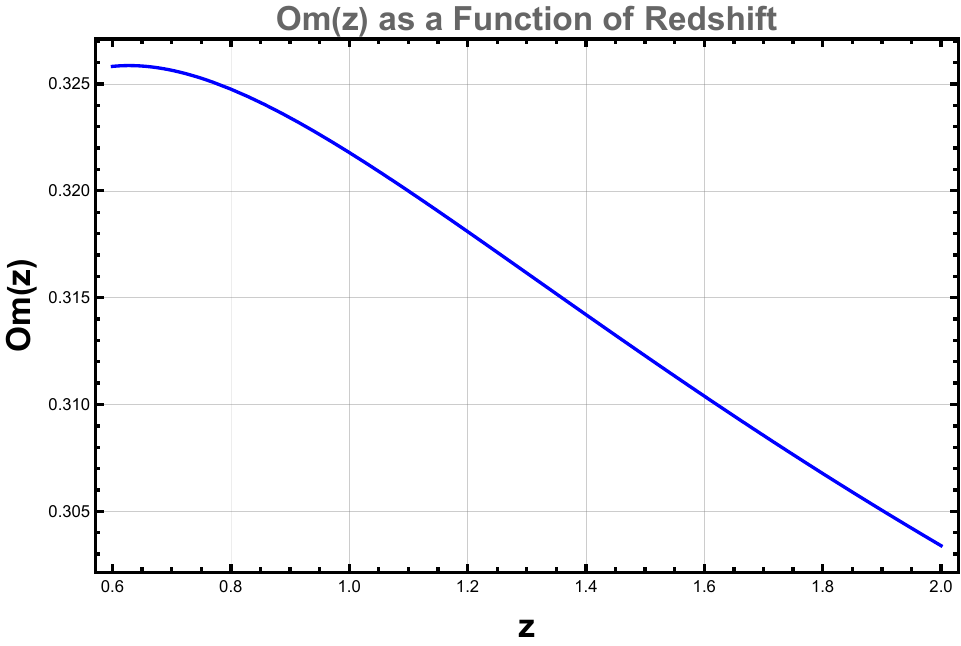}
    \caption{Evolution of the $Om(z)$ parameter as functions of redshift $z$.}
    \label{fig:q_om}
\end{figure}
The analysis of the $Om(z)$ plot from Fig. \ref{fig:q_om} confirms the negative slope, which is compatible with the quintessence behavior of the model.

\section{Conclusion}\label{conclusion}
In this work, we have explored the cosmological viability of a Starobinsky model~\cite{Starobinsky:2007} in the modified symmetric teleparallel gravity model. We have analysed the second-order $f(Q)$ gravity model viability effectively using a combination of recent observational data from cosmic chronometers, PantheonPlus SH0ES, and DESI BAO DR2. We have performed a Markov Chain Monte Carlo (MCMC) technique to constrain the free parameters of the considered model. Further, we have examined the physical behavior of the model using different cosmological parameters. The present density profile with respect to redshift $z$ is shown in Fig. \ref{fig_density}, where it can be seen that energy density lies in the positive regime. The EoS parameter $\omega$ experienced a transition from a decelerated to an accelerated expansion phase (see Fig. \ref{fig_eos}) and places the Universe in the quintessence regime at the present epoch with value $\omega_{0} = -0.73 \pm 0.3$, which is consistent with observational constraints in~\cite{Capozziello:2014}.

The same has been analysed through the behavior of the deceleration parameter displayed in Fig. \ref{fig:q}. The behavior confirms that the model remains physically well-behaved within the constrained parameter space and exhibits a smooth transition from a decelerated to an accelerated expansion phase at redshift $z_{tr} = 0.573$, consistent with the estimate from Planck-CMB 2018 observation within the $\Lambda$CDM model~\cite{Planck:2018vyg}. 
The current deceleration parameter value is obtained as $q_0 = -0.51 \pm 0.6$, confirming the ongoing accelerated expansion of the universe in the dominance of DE. These findings are consistent with the widely accepted value of $q_0$ obtained in ~\cite{Planck:2018vyg, Capozziello:2014}.

One of the significant tools to reconfirm the physical viability of the model is the energy conditions. We have effectively analysed the behavior of the null, dominant, and strong energy conditions (NEC, DEC, and SEC), as displayed in Fig. \ref{fig:energy-conditions}. The NEC remains positive throughout cosmic evolution, signifying that the energy density is non-negative. The DEC is consistently upheld, ensuring that the energy density surpasses the pressure and that energy propagation remains within causal limits. In contrast, the SEC is violated during both the early and later epochs. This violation aligns with the observed accelerated expansion and implies a divergence from traditional matter-dominated models. 
We have demonstrated the alignment of the model towards the quintessence region at present by analysing the $Om(z)$ diagnostic, see Fig. \ref{fig:q_om}. The negative slope of the $Om(z)$ curve confirms that the EoS parameter lies in the quintessence region. 

Overall, our analysis demonstrates that the proposed novel $f(Q)$ gravity model is observationally compatible and theoretically well-founded, providing a compelling geometric framework to account for the current cosmic acceleration.
Future work may extend this analysis to include perturbation-level dynamics and incorporate upcoming high-precision surveys such as the DESI final release, LSST, and Euclid. These will further help in determining whether non-metricity-based modifications of gravity can provide a definitive explanation for cosmic acceleration.

\section*{Acknowledgements}
The authors acknowledge that this research was carried out without any financial support from any funding agency in the public, commercial, or not-for-profit sector.
\bibliographystyle{utphys}
\bibliography{references}
\end{document}